\documentclass[english,aps,prl,showpacs,amsmath,amssymb,twocolumn,noshowpacs,showkeys]{revtex4}
\usepackage{graphicx}
\usepackage{bm}
\begin{document}
\title{A note on entropic uncertainty relations of position and momentum}
\author{Thomas Sch\"urmann}
\affiliation{J\"ulich Supercomputing Centre, J\"ulich Research Centre, 52425 J\"ulich, Germany}

\begin{abstract}
We consider two entropic uncertainty relations of position and momentum recently discussed in literature.
By a suitable rescaling of one of them, we obtain a smooth interpolation of both for high-resolution and low-resolution measurements respectively. Because our interpolation has never been mentioned in literature before, we propose it as a candidate for an improved entropic uncertainty relation of position and momentum. Up to now, the author has neither been able to falsify nor prove the new inequality. In our opinion it is a challenge to do either one.
\end{abstract}

\maketitle

In quantum mechanics position and momentum are complementary observables, satisfying the uncertainty principle of Heisenberg \cite{H27}. As a consequence, their corresponding probability densities cannot both be arbitrarily concentrated. The reason for that phenomenon is the wave-particle duality formalized by the Fourier transform between the position and momentum representation of the state vector $\psi$ in Hilbert space. The state of minimum uncertainty in terms of standard deviations is well known to be satisfied by Gaussian wave functions.

An alternative measure of uncertainty is the entropy of a continuous distribution (differential entropy) \cite{Sh48}. The first statement concerning a lower bound of that measure for the continuous probability densities of position and momentum in quantum mechanics was conjectured by Everett 1957 \cite{E57} and Hirschman \cite{H57}. It was proved in 1975 by Bialynicki-Birula and Mycielski \cite{BM75} and Beckner \cite{B75}. However, the direct physical interpretation of this inequality is not always clear because its scale of measurement sets an arbitrary zero corresponding to a uniform distribution over a unit volume. Any distribution which is more confined than this has less entropy and will be negative. The experimental possibility to check entropy inequalities for continuous variables associated with tomograms (tomographic approach) is discussed in \cite{M1}\cite{M2}.

For practical purposes, the measurement of observables typically involves an apparatus which always introduces a countable partitioning of the spectrum of the operator $A$ into non overlapping subsets (usually called bins). To this partitioning of the spectrum there corresponds the partitioning of the Hilbert space into orthogonal subspaces, each subspace representing one bin. The corresponding projection operator $\hat{P}^A_i$ is associated to the appropriate subspace of the Hilbert space.

For a given pure state of the system, to each projection operator one can assign the probability $p_i^A$ that a measurement will give a value belonging to the $i$-th bin of the partition. The application of this framework to position and momentum requires the partitioning of the spectrum of $\hat{x}$ and $\hat{p}$ into bins which are usually taken all equal to $\delta x$ and $\delta p$.
\begin{figure}[ht]
\includegraphics[width=8.0cm,height=7.0cm]{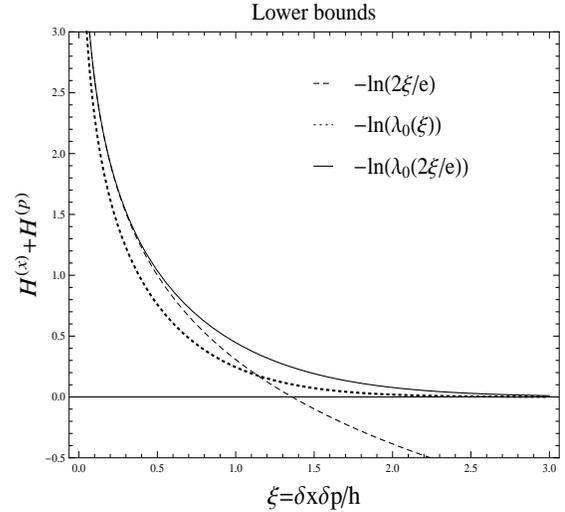}
\caption{The rhs of (\ref{seur}) is negative for $\xi\geq 1.36$ (dashed). The bound in (\ref{lp}) improves (\ref{seur}) for large values of $\xi$ (dotted). The function $-\ln \lambda_0(2\,\xi/e)$ (solid) proceeds slightly above both and asymptotically approaches their envelope.} \label{fig1}
\end{figure}
The probability distributions associated with the finite resolution measurements of position and momentum of a quantum particle in a pure state are defined by
\begin{eqnarray}\label{pos_mom}
q_i=||\hat{P}^x_i\psi||^2\quad\text{and}\quad p_k=||\hat{P}^p_k\psi||^2.
\end{eqnarray}
The indices $i$ and $k$ run from $-\infty$ to $\infty$. From the two probability distributions (\ref{pos_mom}) the Shannon entropies $H^{(x)}$  and $H^{(p)}$ that measure the uncertainty of the  position and the momentum are
\begin{eqnarray}\label{renyixp}
H^{(x)}=-\sum_i q_i\ln q_i, \qquad H^{(p)}=-\sum_k p_k\ln p_k.
\end{eqnarray}
Following Deutsch \cite{D83} and Partovi \cite{P83} the best known lower bound of $H^{(x)}+H^{(p)}$ has been proved by Bialynicki-Birula \cite{BB84}
\begin{eqnarray}\label{seur}
H^{(x)} + H^{(p)}\geq -\ln(2\,\xi/e),
\end{eqnarray}
with $\xi=\delta x\delta p/h$ and the Planck constant $h$. In the quantum regime, when $\xi$ is small this bound gives a meaningful limitation on the sum of the uncertainties in position and momentum. However, this inequality is not sharp since for measurement resolutions $\xi\geq e/2$, this bound becomes trivially satisfied, see Fig.\,\ref{fig1}.\\

On the other hand, Krishna and Parthasarathy \cite{KP02} have generalized a result of Maassen and Uffink \cite{MU88} (Kraus' conjecture \cite{K87}) for pairs of projective measurements of a finite level system. Following the remarks made in \cite{MU88}, it seems possible to extend their result, so as to be applicable also to the case when the Hilbert space is infinitely dimensional, as long as we restrict ourselves to projective measurements of observables. With Theorem 2.1 and Corollary 2.3 of \cite{KP02}, using $||\hat{P}^x_i \hat{P}^p_k||^2=||\hat{P}^x_i \hat{P}^p_k\hat{P}^x_i||$ for all $i,k$ and finally applying a Lemma of \cite{S08}, one gets the lower estimate
\begin{eqnarray}\label{lp}
H^{(x)} + H^{(p)}\geq -\ln \lambda_0(\xi).
\end{eqnarray}
Here $\lambda_0(\xi)$ is the largest eigenvalue of a Fredholm integral equation known in the context of communication theory by Slepian and Pollak \cite{SP61}. The same result has recently been mentioned in \cite{R10}\cite{R11}.\\

\begin{figure}
\includegraphics[width=8.0cm,height=7.0cm]{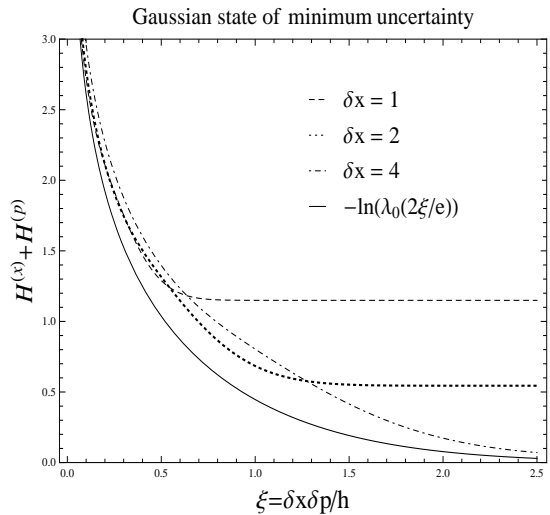}
\caption{Entropy for gaussian states of minimum uncertainty ($\hbar=1, \sigma_x=\sigma_p=1/\sqrt{2}$) and  measurement resolution $\delta x$. The plateaus are saturated at $H^{(x)}$. The rhs of (\ref{conject}) proceeds slightly below the curves in all cases.}\label{fig2}
\end{figure}
Because both bounds above are not tight, one might ask for a nontrivial interpolation which approaches to (\ref{seur}) for small values of $\xi$ and tends to be zero according to (\ref{lp}) for $\xi\to\infty$. Since the function $\lambda_0(\xi)$ monotonically approaches to 1 from below, for $\xi\to\infty$, the right hand side of (\ref{lp}) approaches to zero from above (see Fig.\,\ref{fig1}). On the other hand, in the quantum regime $\xi\to 0$, we have the asymptotic behavior $\lambda_0(\xi)\sim \xi$. Thus, we consider the bound in (\ref{lp}) but replace $\xi$ by the rescaled variable $2\xi/e$ of (\ref{seur}). As a result we obtain the following conjecture
\begin{eqnarray}\label{conject}
H^{(x)} + H^{(p)} \geq -\ln \lambda_0(2\,\xi/e).
\end{eqnarray}
The right hand side proceeds slightly above the envelope of both lower bounds (Fig.\,\ref{fig1}).
We checked (\ref{conject}) for gaussian wave functions in the state of minimum uncertainty by numerical calculations.
In Fig.\,\ref{fig2}, we see the entropy for different values of resolutions $\delta x$. For every fixed $\delta x$ and increasing $\xi$ there is a saturation of the total entropy at the level of $H^{(x)}$. That is because $H^{(p)}$ approaches to zero for increasing values of $\delta p=h\xi/\delta x$.

Further numerical calculations have been considered for single-slit wave functions and for eigenfunctions of the Fourier kernel (spheroidal wave functions). However, in all cases we could not find any example against the inequality (\ref{conject}).



\begin{thebibliography}{99}
\bibitem{H27} W. Heisenberg, Z. Phys. {\bf 43}, 172 (1927).
\bibitem{Sh48} C.\,E. Shannon, Bell Sys. Tech. J. {\bf 27}, 379, 623 (1948).
\bibitem{E57} H. Everett, \emph{The Many World Interpretation of Quantum Mechanics}, (Princeton University Press, Princeton, 1973).
\bibitem{H57} I. Hirschman, Am. J. Math. {\bf 79}, 152 (1957).
\bibitem{BM75} I. Bialynicki-Birula and J. Mycielski, Commun. Math. Phys.  {\bf 44}, 129 (1975).
\bibitem{B75} W. Beckner, Ann. Math. {\bf 102}, 159 (1975).
\bibitem{M1} M.\,A. Man'ko and V.\,I. Man'ko, Found. Phys. {\bf 41}, 330 (2011).
\bibitem{M2} V.\,I. Man'ko, G. Marmo , A. Porzio , S. Solimeno , F. Ventriglia, Phys. Scr. {\bf 83}, No. 4, 045001 (2011).
\bibitem{D83} D. Deutsch,  Phys. Rev. Lett. {\bf 50}, 631 (1983).
\bibitem{P83} M.\,H. Partovi, Phys. Rev. Lett. {\bf 50}, 1883 (1983).
\bibitem{BB84} I. Bialynicki-Birula, Phys. Lett. {\bf 103 A}, 253 (1984).
\bibitem{KP02} M. Krishna and K.\,R. Parthasarathy, Indian J. of Statistics Ser. {\bf A 64}, 842 (2002)
{\tt arxiv:quant-ph/0110025}.
\bibitem{MU88} H. Maassen and J.\,B.\,M. Uffink, Phys. Rev. Lett. {\bf 60}, 1103 (1988).
\bibitem{K87} K. Kraus, Phys. Rev. {\bf D 35}, 3070 (1987).
\bibitem{S08} T. Sch\"{u}rmann, Act. Phys. Pol. {\bf B 39}, 587 (2008) \\ {\tt arxiv:quant-ph/0303005}.
\bibitem{SP61} D. Slepian and H.\,O. Pollak, Bell Syst. Tech. J. {\bf 40}, 43 (1961).
\bibitem{R10} L. Rudnicki (2010) {\tt arxiv:1010.3269 [quant-ph]}.
\bibitem{R11} L. Rudnicki, J. Russ. Laser Res. {\bf 32}, 393 (2011) \\ {\tt arxiv:1108.3828 [quant-ph]}.

\end{thebibliography}
\end{document}